\documentclass[journal,10pt,letterpaper,twoside,twocolumns]{IEEEtran}

\usepackage{graphicx} % Required for inserting images

\usepackage{amsfonts,amsmath,amsthm}
\usepackage[noadjust]{cite}

\usepackage[usenames,dvipsnames]{color}
\usepackage{cancel}
\usepackage{subcaption}
\usepackage[normalem]{ulem}

\title{On the variance of the Least Mean Square squared-error sample curve}
\author{Marcos~H.~Maruo,~\IEEEmembership{Member,~IEEE,} and Jos\'{e}~Carlos~M.~Bermudez,~\IEEEmembership{Senior Member,~IEEE} 
\thanks{This work was partly supported by CNPq under grants No 305377/2009-4 and 473123/2009-6}}%
\date{\today}

\def\sx{x}                   % input signal
\def\sh{h}                   % unknown impulse response
\def\sN{N}                   % adaptive filter length
\def\sNh{\sN}     % unknown impulse response length
\def\sd{d}                   % desired signal
\def\sr{r}                   % near-end or noise signal
\def\se{e}                   % error signal
\def\sJ{J}

\def\opt{\mathrm{opt}}
\def \sJmin{\sJ_{\min}}
\def\ssigmar{\sigma_\sr}
\def\sw{w}
\def\sv{v}

\def\kurtosisr{\psi_\sr}
\def\sJinf{\sJ[\infty]}
\def\enlms{e_{\text{NLMS}}}

\def\vx{{\boldsymbol \sx}}
\def\vxh{{\boldsymbol \sx}}
\def\vh{{\boldsymbol \sh}}

\def\vw{{\boldsymbol \sw}}
\def\vwnlms{{\boldsymbol \sw}_{\text{NLMS}}}
\def\vv{{\boldsymbol \sv}}

\def\mI{{\boldsymbol I}}     % Identity matrix
\def\mO{{\boldsymbol 0}}     % Null matrix

\def\mRxhxh{{\boldsymbol R}_{\mathrm{\sx \sx}}}

\def\mK{{\boldsymbol K}}

\def\mA{{\boldsymbol A}}     

\def \ftr{\mathrm{Tr}}

\begin{document}

\maketitle

\begin{abstract}
	Most studies of adaptive algorithm behavior consider performance measures based on mean values such as the mean-square error. The derived models are useful for understanding the algorithm behavior under different environments and can be used for design.  Nevertheless, from a practical point of view, the adaptive filter user has only one realization of the algorithm to obtain the desired result.  This letter derives a model for the variance of the squared-error sample curve of the least-mean-square (LMS) adaptive algorithm, so that the achievable cancellation level can be predicted based on the properties of the steady-state squared error.  The derived results provide the user with useful design guidelines. 
\end{abstract}

% \fi
\IEEEpeerreviewmaketitle

\section{Introduction}

\IEEEPARstart{A}{daptive} filters encounter many practical applications in online system identification, interference cancellation, prediction and inverse modeling. The least-mean-square (LMS) algorithm is widely used due to its easy implementation and simplicity of design for guaranteed convergence in stationary environments. 

The mean-square error (MSE) is the standard criterion for comparing competing adaptive filtering methods and, despite its shortcomings, it is a) an useful measure of adaptation error energy, b) the dominant metric in the context of optimization problems and c) simple and mathematical tractable~\cite{zhou_wang_mean_2009}.

In general, MSE-based adaptive filter designs are evaluated through learning curves from which one determines convergence rates and steady-state behavior. Analytical models are available to predict the steady-state MSE of the LMS adaptive filter under different practical conditions~\cite{bershad_analysis_1986, haykin_adaptive_1991, hyun-chool_shin_transient_2003, manolakis_statistical_2005, sayed_adaptive_2008, de_almeida_stochastic_2009}. Such predictions are very important for design purposes, as they inform the designer about the expected level of estimation error cancellation that should be expected for a given set of parameter values and signal properties.  Nevertheless, the success of an adaptive filter implementation in a real time environment is determined by a single run of the algorithm. 

Even considering its importance for a practical design, to the best of our knowledge very few works have addressed the behavior of sample learning curves of adaptive algorithms. The work \cite{nascimento_on_2000, nascimento_are_1998} studied the validity of the ensemble-average learning curves for evaluating the performance of adaptive filters.  The authors showed that approximating transient of the LMS MSE learning curves by performing several repeated experiments and by averaging the resulting squared-error curves can be misleading for large step sizes. The work in \cite{Solo_limiting_1989} derived realization-wise results for noise misadjustment and lag-misadjustment of the LMS algorithm after convergence. It was shown that the classical lag-misadjustment expression misrepresented the effects of noise-misadjustment and noise variance.   

In this paper we look at the steady-state cancellation error considering not only the MSE, but also the variance of the squared-error curve. This is an important study for the designer, as the actual level of the estimation error is determined by the specific sample function of the squared error generated in that single realization. Thus, for an estimation error $e[n]$ the actual achievable level of estimation error cancellation is determined by the properties of the random variable $e^2[n]$ in steady-state.  In the case of stationary environments, these properties include the mean and the variance of the squared error, as the latter may end up being much higher than the MSE, in which most designs are usually based. 
 
% 
%The paper is organized as follows\cblue{. Section \ref{sec:perfsurface} describes the performance surface of the optimization problem and the optimal solution. Section \ref{sec:LMS} describes the adaptive algorithm. In Section \ref{sec:squarederror} an statistical analysis of the steady squared errror is performed and the results are used to predict the steady-state squared error behavior of the LMS and NLMS algorithms. In Section \ref{sec:results}, the derived model is then used to predict the filter behavior with different noise distributions. }
%\cblue{In this paper, plain letters denote scalars, lowercase boldface letters denote column vectors and uppercase boldface letters denote matrices.}

The paper is organized as follows. Section \ref{sec:perfsurface} describes the performance surface of the optimization problem and the optimal solution. Section \ref{sec:LMS} describes the adaptive algorithm. Section \ref{sec:squarederror} derives an analytical model for the variance of the squared error for the LMS and NLMS algorithms. Section \ref{sec:results} applies the derived model to predict the filter behavior for different noise distributions. 
In this paper, plain letters denote scalars, lowercase boldface letters denote column vectors and uppercase boldface letters denote matrices.

\section{Performance surface}
\label{sec:perfsurface}
Consider a linear adaptive filter with input vector 
$\vxh[n] = [\sx[n]\thickspace \sx[n-1] \thickspace \cdots \thickspace \sx[n-\sNh]]^\top$,
coefficient vector 
$\vw[n] = [ \sw_0[n] \thickspace \sw_1[n] \thickspace \cdots \thickspace \sw_0[\sNh-1]]^\top$
and the desired output signal 
\begin{align}
\sd[n] = \sr[n] + \vxh^\top[n] \vh 
\label{eqn:d}
\end{align}
where $\vh = [\sh_0 \thickspace \sh_1 \thickspace \cdots \thickspace \sh_{\sNh-1}]^\top$ is the unknown impulse response and $\sr[n]$ is a white noise statistically independent of $\vxh[n]$.
The MSE performance surface is defined by
\begin{align}
J(\vw) = E\{\se^2[n]|\vw[n] = \vw\} = \ssigmar^2 + E\{ [\vxh^\top[n] (\vh-\vw)]^2 \}
\label{eqn:MSE2}
\end{align}
where $\se[n]$ denotes the approximation error 
\begin{align}
 \se[n] = \sd[n] - \vxh^\top[n]\vw.
\label{eqn:e}
\end{align}

It is easy to see that the minimum of~\eqref{eqn:MSE2} is achieved for $\vw_{\opt}=\vh$, where $\vw_{\opt}$ denotes the optimal solution, so that $\sJmin = J(\vw_{\opt}) = E\{\sr^2[n]\} =  \ssigmar^2$. 
	
The performance of the adaptive algorithm is usually studied through the properties of the weight-error vector
\begin{align}
\vv[n] = \vw[n] - \vh.
\label{eqn:v}
\end{align}
In the following we study the behavior of the squared-error variance as a function of $\vv[n]$, the input signal statistical properties, and the adaptation step-size. 

\section{The LMS algorithm}
\label{sec:LMS}

The LMS weight update equation is \cite{sayed_adaptive_2008}

\begin{equation}
	\vw[n] = \vw[n-1] + \mu e[n]\vx[n]
\label{eqn:LMS_w}
\end{equation}
where $\mu$ is the learning step-size. Using \eqref{eqn:v} in \eqref{eqn:LMS_w} yields
\begin{equation}
	\vv[n] = \vv[n-1] + \mu e[n]\vx[n].¨
	\label{eqn:LMS_v}
\end{equation}

\section{Squared-error variance}
\label{sec:squarederror}
Define $J[n] = J(\vw[n])$. Then, the squared error variance is given by
\begin{align}
\sigma_{\se^2}^2[n] = E\{ \se^4[n] \} - \sJ^2[n]
\label{eqn:MSEvariance}
\end{align}
where $e[n]$ can be written as a function of the weight-error vector using \eqref{eqn:v} as 
\begin{equation}
\se[n] =  \sr[n] - \vxh^\top[n] \vv[n].
\label{eqn:errorfunctionofv}
\end{equation}
Then, the non-central fourth order moment of \eqref{eqn:errorfunctionofv} is given by
\begin{align}
	E\{\se^4[n]\} & = E\{ (\sr[n] + \vv^\top[n] \vxh[n])^4 \}\nonumber\\
	&  =  E\{ \sr^4[n]\} +4 E\{ \sr^3[n] \vxh^\top[n] \vv[n] \}\nonumber\\
	& + 6 E\{ \sr^2[n] (\vv^\top[n] \vxh[n] )^2 \}  \nonumber\\
	& + 4 E\{ \sr[n] (\vv^\top[n] \vxh[n] )^3 \} 
	+ E\{ (\vv^\top[n] \vxh[n] )^4 \}.
	\label{eqn:e4}
\end{align}

\subsection{Simplifying Assumptions}
For the analysis we consider the following typical simplifying assumptions for mathematical tractability~\cite{haykin_adaptive_1991}:
\begin{description}
 \item[A1] $\vxh[n]$ is a zero-mean Gaussian vector;
 \item[A2] $\sr[n]$ is zero-mean white and statistically independent of any other signal;
 \item[A3] The input autocorrelation matrix $\mRxhxh = E[\vxh[n]\vxh^{\top}[n]]$ is a positive-definite matrix;    
 \item[A4] The statistical dependence between $\vxh[n]\vxh^T[n]$ and $\vw[n]$ can be neglected.
\end{description}

Though assumption A4 is not strictly valid, it has been shown that its use leads to analytical models that accurately predict the behavior of adaptive filters~\cite{haykin_adaptive_1991, manolakis_statistical_2005, sayed_adaptive_2008}.  

We now study each term of \eqref{eqn:e4}.

\subsection{Term $E\{ \sr^4[n]\}$}
The first term of~\eqref{eqn:e4} is determined straightforward from
\begin{align}
E\{ \sr^4[n]\} = E^2\{ \sr^2[n]\} \kurtosisr = \sJmin^2\kurtosisr
\label{eqn:exodia1}
\end{align}
where $\kurtosisr = \frac{E\{\sr^4[n]\}}{\ssigmar^4}$ is the kurtosis of the noise distribution. In particular, for Gaussian $\sr[n]$ we have $E\{ \sr^4[n]\} = 3\sJmin^2$.

\subsection{Term $E\{ \sr^3[n] \vxh^\top[n] \vv[n] \}$}
Using A2 and A4 on the second term of~\eqref{eqn:e4} yields
\begin{align}
E\{ \sr^3[n] \vxh^\top[n] \vv[n] \} & =  E\{ \sr^3[n]\} E\{\vxh^\top[n]\}E\{ \vv[n] \}
% \end{align}
% where $E\{ \sr^3[n] \vxh[n]\}^\top = E\{ \sr^3[n]\} E\{\vxh^\top[n]\} = $ yields
% \begin{align}
%[E\{\sr^3[n] \vxh[n]\}]_i = E\{ \sr^3[n] \sx[n-i+1] \} %= 3 E\{ \sr^2[n] \} E\{ \sr[n] \sx[n-i+1] \}
=0
\label{eqn:exodia2}
\end{align}
where $E\{ \vx[n] \} = \mO$ from A1.
% Therefore, substituting~\eqref{eqn:Pv} in~\eqref{eqn:gaussmomentfac2ndterme4} yields
% \begin{align}
% E\{ \sr^3[n] \vxh^\top[n] \vv[n] \} =  E\{ \sr^2[n] E\{\vxh^\top[n] \sr[n]\} \mP^\top E\{ \vv[n] \} = E\{ \sr^2[n] (\mP E\{\vxh[n] \sr[n]\})^\top E\{ \vv[n] \} = 0
% \end{align}
% where $\mP E\{\vxh[n] \sr[n]\} = \mO_{\sNh \times 1}$ from~\eqref{eqn:Pxhe0}

\subsection{Term $E\{ \sr^2[n] (\vv^\top[n] \vxh[n] )^2 \}$}
Using A4 on the third term of~\eqref{eqn:e4} yields
\begin{equation}
\begin{split}
E\{ &\sr^2[n] (\vv^\top[n] \vxh[n] )^2 \} = \\%& = E\{ \sr^2[n] \sum_{ k=1 }^{\sNh} [\vv[n]]_k \sx[n -k] \sum_{ j=1 }^{\sNh} [\vv[n]]_j \sx[n -j] \}\\
 &\sum_{ k=1 }^{\sNh} \sum_{ j=1 }^{\sNh}  E\{ \sr^2[n] \sx[n -k] \sx[n -j] \}  E\{ [\vv[n]]_k [\vv[n]]_j \}
 \end{split}
\label{eqn:e43rdtermexpanded}
\end{equation}
where $[\vv[n]]_k$ denotes the $k$th component of $\vv[n]$. Then, using A2 in $E\{ \sr^2[n] \sx[n -k] \sx[n -j] \}$ yields
\begin{align}
E\{ \sr^2[n] \sx[n -k] \sx[n -j] \} = \sJmin E\{ \sx[n -k] \sx[n -j] \}. % + 2 E\{ \sr[n] \sx[n -k]\} E \{ \sr[n] \sx[n -j]\}  
\label{eqn:e43rdtermgaussmomentfact}
\end{align}
%where $E\{ \sr[n] \sx[n-i+1] \} = E\{ \sr[n]\}E\{ \sx[n-i+1] \} = 0$ from A1.
Substituting~\eqref{eqn:e43rdtermgaussmomentfact} %,~\eqref{eqn:Pv} and~\eqref{eqn:Pxhe0} 
in \eqref{eqn:e43rdtermexpanded} and defining $\mK[n]= E\{\vv[n]\vv^\top[n]\}$ yields
\begin{align}
\begin{split}
E\{ &\sr^2[n] (\vv^\top[n] \vxh[n] )^2 \} \\
& = \sJmin \sum_{ k=1 }^{\sNh} \sum_{ j=1 }^{\sNh} E\{ \sx[n -k] \sx[n -j] \} E\{ [\vv[n]]_k [\vv[n]]_j \} \\
%& + 2 \sum_{ k=1 }^{\sNh} \sum_{ j=1 }^{\sNh} E\{ \sr[n] \sx[n -k]\} E \{ \sr[n] \sx[n -j]\} E\{ [\vv[n]]_k [\vv[n]]_j \} \nonumber\\
%& = \sJmin \ftr(  \mK[n]  \mP \mRxhxh \mP^\top ) + E\{ \sr[n] \vxh[n] \} ^\top \mP^\top \mK[n] \mP  E\{ \sr[n] \vxh[n] \} \nonumber\\
& = \sJmin \ftr(  \mK[n]  \mRxhxh ). 
\end{split}
\label{eqn:exodia3}
\end{align}
%where~\eqref{eqn:Pxhe0} was used.

\subsection{Term $E\{ \sr[n] (\vv^\top[n] \vxh[n] )^3 \}$}
Using A2
% Using A2 and A4 on the fourth term of~\eqref{eqn:e4} yields
% \begin{align}
% E\{ \sr[n] (\vv^\top[n] \vxh[n] )^3 \} % & = E\{ \sr[n] \sum_{i=1}^{\sNh} [\vv[n]]_i \sx[n-i]  \sum_{j=1}^{\sNh} [\vv[n]]_j \sx[n-j] \sum_{k=1}^{\sNh} [\vv[n]]_k \sx[n-k] \}\\
% & = \sum_{i=1}^{\sNh} \sum_{j=1}^{\sNh} \sum_{k=1}^{\sNh} E\{ \sr[n] \}E\{\sx[n-i]  \nonumber\\
% & \quad \sx[n-j]\sx[n-k] \}  E\{ [\vv[n]]_i [\vv[n]]_j \nonumber\\
% & \quad [\vv[n]]_k \}
% \label{eqn:e44thtermexpanded}
% \end{align}
% where, using A1 on \eqref{eqn:e44thtermexpanded} yields
%the Gaussian moment factoring theorem on $E\{ \sr[n] \sx[n-i] \sx[n-j] \sx[n-k] \}$ yields
%\begin{align}
%E\{ \sr[n] \sx[n-i] \sx[n-j] \sx[n-k] \} %& = E\{ \sr[n] \sx[n-i] \}E\{\sx[n-j] \sx[n-k] \} + E\{ \sr[n] \sx[n-j] \}E\{\sx[n-i] \sx[n-k] \}\nonumber\\
%%& + E\{ \sr[n] \sx[n-k] \}E\{\sx[n-j] \sx[n-j] \}  
%= 0
%\label{eqn:e44thtermgaussmomentfact}
%\end{align}
%where $E\{ \sr[n] \sx[n-i+1] \} = E\{ \sr[n]\}E\{ \sx[n-i+1] \} = 0$ from A1. 
%Substituting \eqref{eqn:e44thtermgaussmomentfact} %and~\eqref{eqn:Pv} 
%in \eqref{eqn:e44thtermexpanded} yields
\begin{align}
E\{ \sr[n] (\vv^\top[n] \vxh[n] )^3 \} = E\{ \sr[n]\}E\{(\vv^\top[n] \vxh[n] )^3 \}
= 0.
\label{eqn:exodia4}
\end{align}
%where $\mP E\{\vxh[n]\sr[n]\} = \mO_{\sNh \times 1}$ from~\eqref{eqn:Pxhe0}

\subsection{Term $E\{ (\vv^\top[n] \vxh[n] )^4 \}$}
Using A4 in the fifth term of~\eqref{eqn:e4} yields
\begin{equation}
\begin{split}
E\{ &(\vv^\top[n] \vxh[n] )^4 \} \\%& = E\{ \sum_{i=1}^{\sNh} [\vv[n]]_i \sx[n-i] \sum_{j=1}^{\sNh} [\vv[n]]_j \sx[n-j] \sum_{k=1}^{\sNh} [\vv[n]]_k \sx[n-k] \sum_{\ell=1}^{\sNh} [\vv[n]]_{\ell} \sx[n-\ell] \}\nonumber\\ 
& = \sum_{i=1}^{\sNh} \sum_{j=1}^{\sNh} \sum_{k=1}^{\sNh} \sum_{\ell=1}^{\sNh} E\{ \sx[n-i]   \sx[n-j] \sx[n-k]   \sx[n-\ell] \}\\
&\hspace{18ex}\times E\{ [\vv[n]]_i [\vv[n]]_j [\vv[n]]_k [\vv[n]]_{\ell}\}.
\end{split}
\label{eqn:e45thtermexpanded}
\end{equation}
Using the Gaussian moment factoring theorem yields
\begin{align} 
\begin{split}
E\{ &\sx[n-i]  \sx[n-j]   \sx[n-k]   \sx[n-\ell] \} =  \\
&E\{ \sx[n-i]   \sx[n-j] \}  E\{ \sx[n-k]   \sx[n-\ell] \}  \\
& + E\{ \sx[n-i]   \sx[n-k] \}  E\{ \sx[n-j]   \sx[n-\ell] \} \\
& + E\{ \sx[n-i]   \sx[n-\ell] \}  E\{ \sx[n-j]   \sx[n-k] \}
\end{split}
\label{eqn:pedaco1navemae}
\end{align}

%\begin{align}
%E\{ \sx[n-i]   \sx[n-j]   \sx[n-k]   \sx[n-\ell] \} & = E\{ \sx[n-i]  % \sx[n-j] \}  E\{ \sx[n-k]   \sx[n-\ell] \} \nonumber \\
%& + E\{ \sx[n-i]   \sx[n-k] \}  E\{ \sx[n-j]   \sx[n-\ell] %\}\nonumber \\
%& + E\{ \sx[n-i]   \sx[n-\ell] \}  E\{ \sx[n-j]   \sx[n-k] \}
%\label{eqn:pedaco1navemae}
%\end{align}

The evaluation of the last term in \eqref{eqn:e45thtermexpanded} one needs the statistical properties of $\vv[n]$. The distribution of $\vv[n]$ is unknown and depends on the filter initialization strategy, as well on the noise and input signal distributions. However, defining $n_0$ as the starting moment of the converged steady-state MSE and after a large number of iterations, we have from \eqref{eqn:d}, \eqref{eqn:e} and \eqref{eqn:LMS_v}, that
\begin{equation}
	\vv[n] = \vv[n-n_0] + \mu \sum_{k=0}^{n_0 - 1} r[n-k]\vx[n-k].
\label{eq:ss_vn}
\end{equation} 
Also as $\lim_{n\to\infty}E\{\vw[n]\} = \vw_{\opt} = \vh$, then $\lim_{n\to\infty}E\{\vv[n]\}= \boldsymbol{0}$. Then, assuming that $v[n]$ is already stationary for $n \ge n_0$, the steady-state $\vv[n]$ is composed by a large sum of stationary random vectors.  Hence, it is reasonable to approximate the distribution of $\lim_{n \rightarrow \infty}\vv[n]$ by a zero-mean Gaussian. Such an approximation has been successfully used in the analysis of the sign algorithm, which includes even a nonlinearity in the weight updating term~\cite{koike_convergence_1995}. Then, applying the Gaussian moment factoring theorem yields

\begin{align}
\begin{split}
E\{ [\vv[n]]_i &[\vv[n]]_j [\vv[n]]_k [\vv[n]]_{\ell}\}  = \\
&E\{ [\vv[n]]_i [\vv[n]]_j \}E\{[\vv[n]]_k [\vv[n]]_{\ell}\}\\
& +  E\{ [\vv[n]]_i [\vv[n]]_k \}E\{[\vv[n]]_j [\vv[n]]_{\ell}\}\\
& + E\{ [\vv[n]]_i [\vv[n]]_{\ell} \}E\{[\vv[n]]_j [\vv[n]]_k\}.
\end{split}
\label{eqn:pedaco2navemae}
\end{align}
%\cred{Perguntar para o Bermudez se $\lim_{n \rightarrow \infty}\mK[n]$ tem alguma propriedade interessante (diagonal? relacionada a $\mRxhxh$ por uma transformação de similaridade?) em ambos os casos deve ser possível usar o resultado do Eweda e não o teorema da fatoração dos momentos Gaussianos}

Substituting \eqref{eqn:pedaco1navemae} and \eqref{eqn:pedaco2navemae} in  \eqref{eqn:e45thtermexpanded} yields
\begin{align}
E\{ (\vv^\top[n] \vxh[n] )^4 \} %& = \sum_{i=1}^{\sNh} \sum_{j=1}^{\sNh} \sum_{k=1}^{\sNh} \sum_{\ell=1}^{\sNh} ( E\{ \sx[n-i]   \sx[n-j] \}  E\{ \sx[n-k]   \sx[n-\ell] \}  E\{ [\vv[n]]_i [\vv[n]]_j \}E\{[\vv[n]]_k [\vv[n]]_{\ell}\} .\nonumber\\
& = 3 \ftr^2(\mRxhxh \mK[n]) \nonumber\\
&\hspace{2ex} + 6 \ftr(\mRxhxh \mK[n]  \mRxhxh \mK[n])
\label{eqn:exodia5}
\end{align}
where $\ftr(\mA)$ denotes the trace of matrix $\mA$. 

Finally, subsituting~\eqref{eqn:exodia1},\eqref{eqn:exodia2},\eqref{eqn:exodia3},\eqref{eqn:exodia4} and~\eqref{eqn:exodia5} in \eqref{eqn:e4} for $n \rightarrow \infty$ yields
\begin{align}
%\lim_{n \rightarrow \infty}E\{\se^4[n]\} 
E\{\se^4[\infty]\} 
& = %\lim_{n \rightarrow \infty} 
\sJmin^2\kurtosisr + 6 \sJmin \ftr(\mK[\infty] \mRxhxh ) \nonumber\\
&\hspace{2ex} + 3 \ftr^2(\mK[\infty] \mRxhxh )\nonumber\\
&\hspace{2ex} + 6 \ftr(\mRxhxh \mK[\infty]  \mRxhxh \mK[\infty]).
\label{eqn:exodia}
\end{align}
Hence, substituting~\eqref{eqn:exodia} in~\eqref{eqn:MSEvariance} yields
\begin{align}
%\lim_{n \rightarrow \infty} \sigma_{\se^2}^2[n] %& = E\{ (\se^4[n] -E^2\{\se^2[n]\}) \} = E\{ \se^4[n]\} - E^2\{ \se^2[n] \} \nonumber\\
\sigma_{\se^2}^2[\infty]
& = (\kurtosisr-1)\sJmin^2 + 4 \sJmin \ftr(\mK[\infty]\mRxhxh ) \nonumber\\
&\hspace{2ex}+ 2 \ftr^2(\mK[\infty] \mRxhxh ) \nonumber\\
&\hspace{2ex}+ 6 \ftr(\mRxhxh \mK[\infty] \mRxhxh \mK[\infty]).
%& = 2 \sJinf^2 + 6 \lim_{n \rightarrow \infty}  \ftr(\mRxhxh \mK[n] \mRxhxh \mK[n]) %= 2 [\sJmin (1 + \frac{\frac{1}{2} \mu \ftr(\mRxhxh)}{1 - \frac{1}{2} \mu \ftr(\mRxhxh)}) ]^2 + 6 \lim_{n \rightarrow \infty}  \ftr(\mRxhxh \mK[n] \mRxhxh \mK[n]) 
\label{eqn:steady-statee2variance}
\end{align}
For Gaussian noise, $\kurtosisr = 3$ and
\begin{align}
%\lim_{n \rightarrow \infty} \sigma_{\se^2}^2[n] 
\sigma_{\se^2}^2[\infty] 
& = 2 \sJinf^2 + 6 %\lim_{n \rightarrow \infty}  
\ftr(\mRxhxh \mK[\infty] \mRxhxh \mK[\infty])
\label{eqn:steady-statee2variancegaussian}
\end{align}

% \subsection{LMS}
% For the LMS, we have~\cite{}
% \begin{align}
% \mK[\infty] = \frac{\mu\sJmin}{2 - \mu \ftr(\mRxhxh)}  \mI_{\sNh}
% \label{eqn:KinfLMS}
% \end{align}
% % Then, in steady-state we have
% % \begin{align}
% % \lim_{n \rightarrow \infty}  \ftr(\mRxhxh \mK[n] \mRxhxh \mK[n]) = \frac{\mu^2\sJmin^2}{(2 - \mu \ftr(\mRxhxh))^2} \ftr (\mRxhxh^2)
% % \end{align}
% and
% \begin{align}
% \sJ[\infty] = \sJmin(1 + \frac{\mu \ftr(\mRxhxh) }{2 - \mu \ftr(\mRxhxh)}) = \frac{2 \sJmin}{2 -\mu \ftr(\mRxhxh)}
% \label{eqn:JinfLMS}
% \end{align}
% % Therefore, from \eqref{eqn:JinfLMS} we have
% % \begin{align}
% % \sJinf^2 & = \frac{4 \sJmin^2}{4 - 4\mu \ftr(\mRxhxh) + \mu^2 \ftr^2(\mRxhxh)}
% % \label{eqn:JinfLMSsquared}
% % \end{align}

% Substituting \eqref{eqn:KinfLMS} and \eqref{eqn:JinfLMS} in \eqref{eqn:steady-statee2variance} yields
% \begin{align}
% \lim_{n \rightarrow \infty} \sigma_{\se^2}^2[n] = \sJmin^2 (\frac{2 + \frac{3}{2} \mu^2\ftr (\mRxhxh^2)}{1 - \mu \ftr(\mRxhxh)(1 - \frac{\mu}{4} \ftr(\mRxhxh) )} )
% \end{align}

\subsection{LMS Algorithm}

It has been shown in \cite{butterweck_a_1995} that the steady-state weight-error covariance matrix of the LMS algorithm for small step-sizes can be approximated by the diagonal matrix 
\begin{align}
	\mK[\infty] = \frac{\mu\sJmin}{2}  \mI.
	\label{eqn:KinfLMS2}
\end{align}
where $\mI$ is the identity matrix. Also, the steady-state MSE is given by \cite{sayed_adaptive_2008} 
\begin{align}
\sJ[\infty] = \sJmin\left(1 + \frac{\mu}{2} \ftr(\mRxhxh) \right).
\label{eqn:JinfLMS2}
\end{align}

Substituting \eqref{eqn:KinfLMS2} %and \eqref{eqn:JinfLMS2} in \eqref{eqn:steady-statee2variancegaussian} 
in \eqref{eqn:steady-statee2variance} 
yields the expression for the variance of the steady-state squared error:
\begin{align}
%\lim_{n \rightarrow \infty} \sigma_{\se^2}^2[n]  
% \sigma_{\se^2}^2[\infty] = &  \sJmin^2 \left[ 2 + 2\mu\ftr(\mRxhxh) + \frac{\mu^2}{2} (\ftr^2( \mRxhxh) \right.\nonumber\\
% & \left.+  3 \ftr(\mRxhxh^2 )) \right].
\sigma_{\se^2}^2[\infty] = & \sJmin^2 \Big[(\psi_r-1) + 2 \mu \ftr(\mRxhxh) + \frac{\mu^2}{2}(\ftr^2(\mRxhxh) \nonumber\\
& + 3\ftr(\mRxhxh^2))\Big].
\label{eqn:maisfacildeinterpretar}
\end{align}

We note that the derivative of \eqref{eqn:maisfacildeinterpretar} with respect to the step-size $\mu$ is positive, since $\mRxhxh$ and $\mRxhxh^2$ are positive definite matrices and $\sJmin>0$. This result confirms the expected property that the variance of the steady-state squared error should increase with $\mu$. More than that, \eqref{eqn:maisfacildeinterpretar} now provides a closed expression to predict such variance as a function of the design parameters.

\subsection{NLMS Algorithm}

The analysis of the LMS ensemble curves can be also useful for the design of the Normalized Least Mean-Square (NLMS) algorithm for the practical case of long filters, as in this case NLMS behaves approximately as LMS with a constant normalized step-size.

The NLMS weight update is given by
\begin{equation}
	\vwnlms[n] = \vwnlms[n-1] + \frac{\beta}{\vx^\top[n]\vx[n]} \enlms[n]\vx[n]
\end{equation}

For a large number of coefficients and stationary input, the term $\vx^\top[n]\vx[n]$ can be approximated by $\sN \sigma_\sx^2$~\cite{bershad_analysis_1986, costa_improved_2002}. Under this approximation, \eqref{eqn:maisfacildeinterpretar} can be used with the step-size $\mu$ replaced with $\beta/(N\sigma_\sx^2)$ to predict the behavior of the NLMS squared error.

\section{Results}
\label{sec:results}

The simulations in this section were performed using an AR1 input signal with $0.5$ correlation coefficient and unitary variance. The additive noise variance was $\ssigmar^2\times 10^{-6}$, regardless of the statistical distribution.
The LMS and NLMS step-sizes were $\mu=2/(30\sigma_{\sx}^2 N)$ and $\beta=0.1$ respectively.

\subsection{Gaussian noise}
For a Gaussian noise, the converged error $\se[\infty]$ is also Gaussian. Then, $\se^2[\infty]/J[\infty]$ is a chi-square random variable with 1 degree of freedom. In this case, the 95\% and 99.7\% chi-square confidence intervals with equal areas around the median are shown in Table~\ref{tbl:chi-squared}. Hence, the steady-state confidence intervals for $\se^2[n]$ are obtained by multiplying the values in Table~\ref{tbl:chi-squared} by  $\sJ[\infty]$ from ~\eqref{eqn:JinfLMS2} (with $\mu =\beta/(\sN\sigma_\sx^2)$ for NLMS). Then, the upper limit of the 99.7\% confidence interval is approximately 10.4 dB above $J[\infty]$.

\begin{table}[htp]
\centering
\caption{Confidence intervals for a chi-squared random variable with 1 degree of freedom}
\label{tbl:chi-squared}
\begin{tabular}{|c|l|l|}
\hline confidence & minimum & maximum\\
\hline 95\% & 0.0009820691171752583 & 5.023886187314888\\
\hline 99.7\% & 3.5342958990342576e-06 & 10.078615499494532 \\
\hline
\end{tabular}
\end{table}

\begin{figure}[htp]
     \begin{subfigure}[b]{.22\textwidth}
        \includegraphics[width=\textwidth]{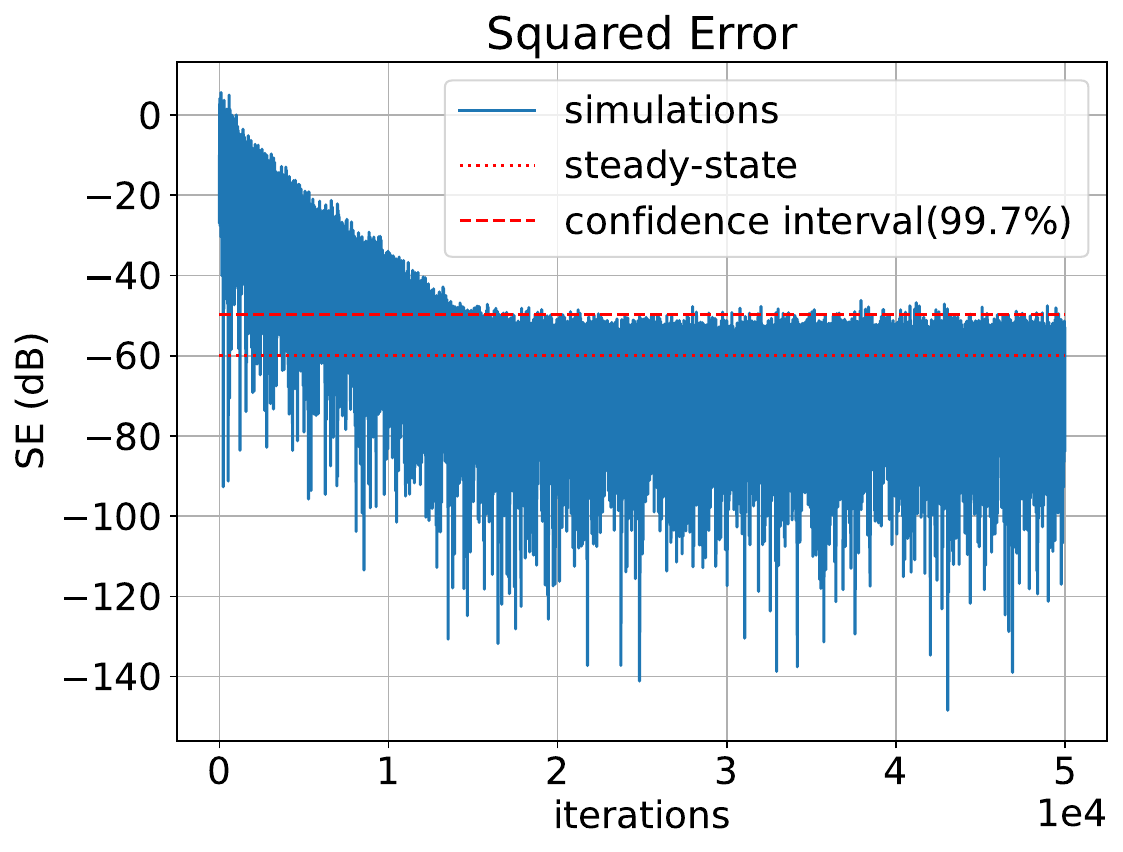}
        \caption{LMS (Gaussian noise)}
        \label{fig:LMSgaussian}
     \end{subfigure}
%     \hfill   
     \begin{subfigure}[b]{.22\textwidth}
        \includegraphics[width=\textwidth]{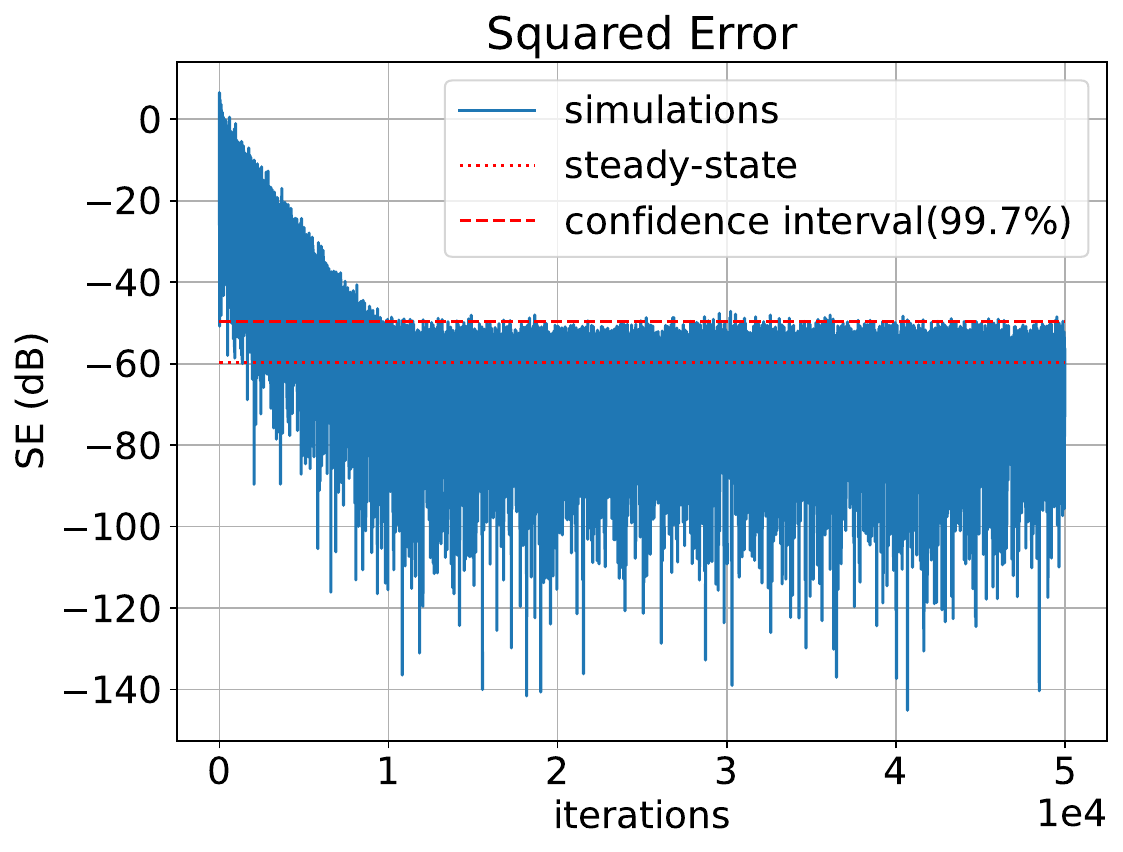}
        \caption{NLMS (Gaussian noise)}
        \label{fig:NLMSgaussian}
     \end{subfigure}
\caption{Sample curves for $\se^2[n]$ and predicted 99.7\% confidence intervals for Gaussian noise.}
\label{fig:gaussian}
\end{figure}

The unknown system $\vh$ was the m1 response defined in~\cite{ITU-TG.168}. The additive noise was WGN ($\kurtosisr=3.0$). Figure~\ref{fig:gaussian} shows the evolution of a single run of the squared error (SE) $\se^2[n]$ for LMS and NLMS. The horizontal lines show the theoretical 99.7\% confidence intervals. These results clearly show that the theoretical upper limit provides a useful information for design purposes. 

\subsection{Non-Gaussian noise}
\begin{figure}[htp]
     \begin{subfigure}[b]{.22\textwidth}
        \includegraphics[width=\textwidth]{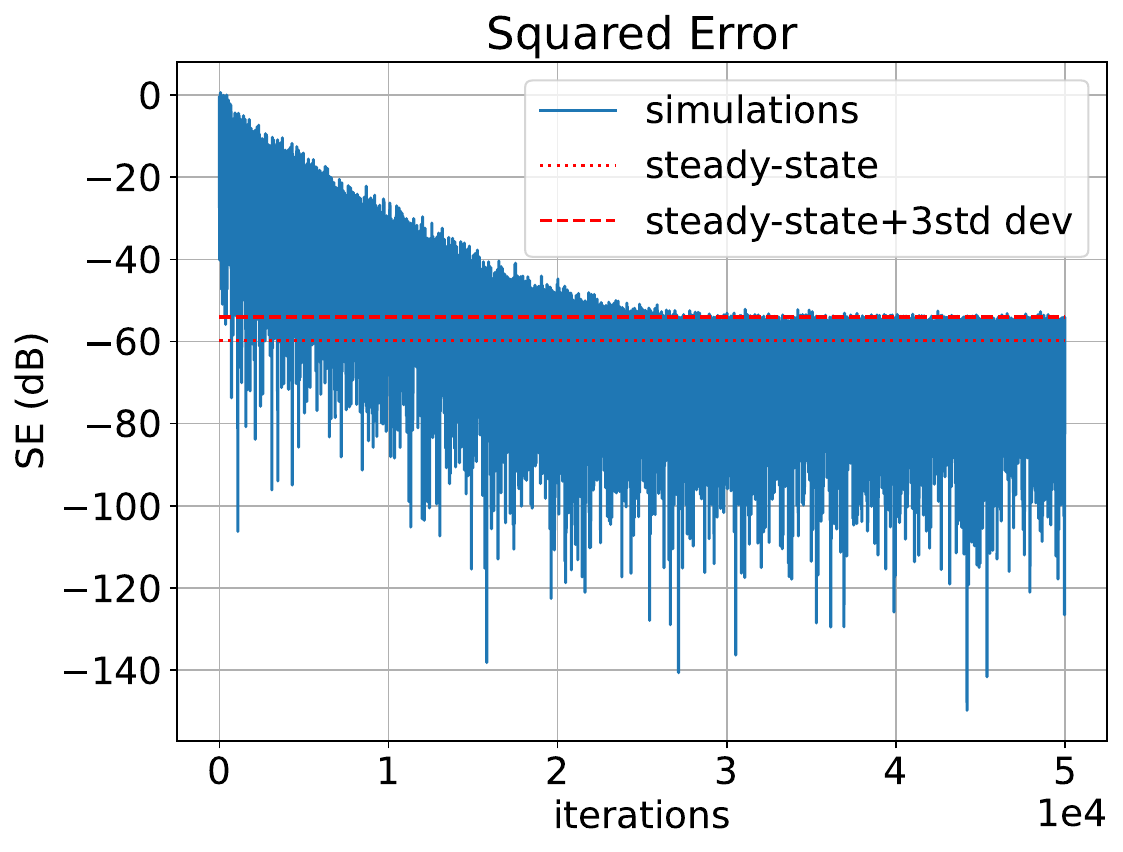}
        \caption{LMS (uniform noise)}
        \label{fig:LMSuniform}
     \end{subfigure}
     \begin{subfigure}[b]{.22\textwidth}
        \includegraphics[width=\textwidth]{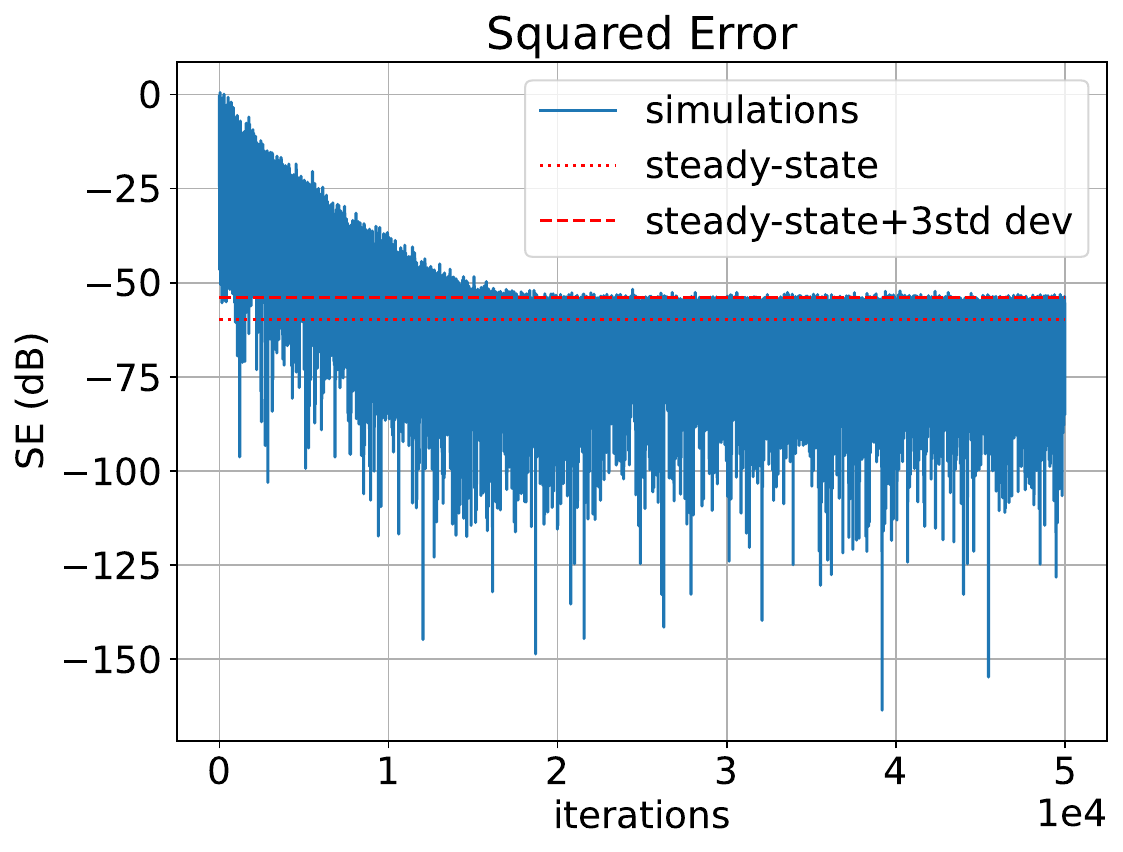}
        \caption{NLMS (uniform noise)}
        \label{fig:NLMSuniform}
     \end{subfigure}
     \hfill   
     \begin{subfigure}[b]{.22\textwidth}
        \includegraphics[width=\textwidth]{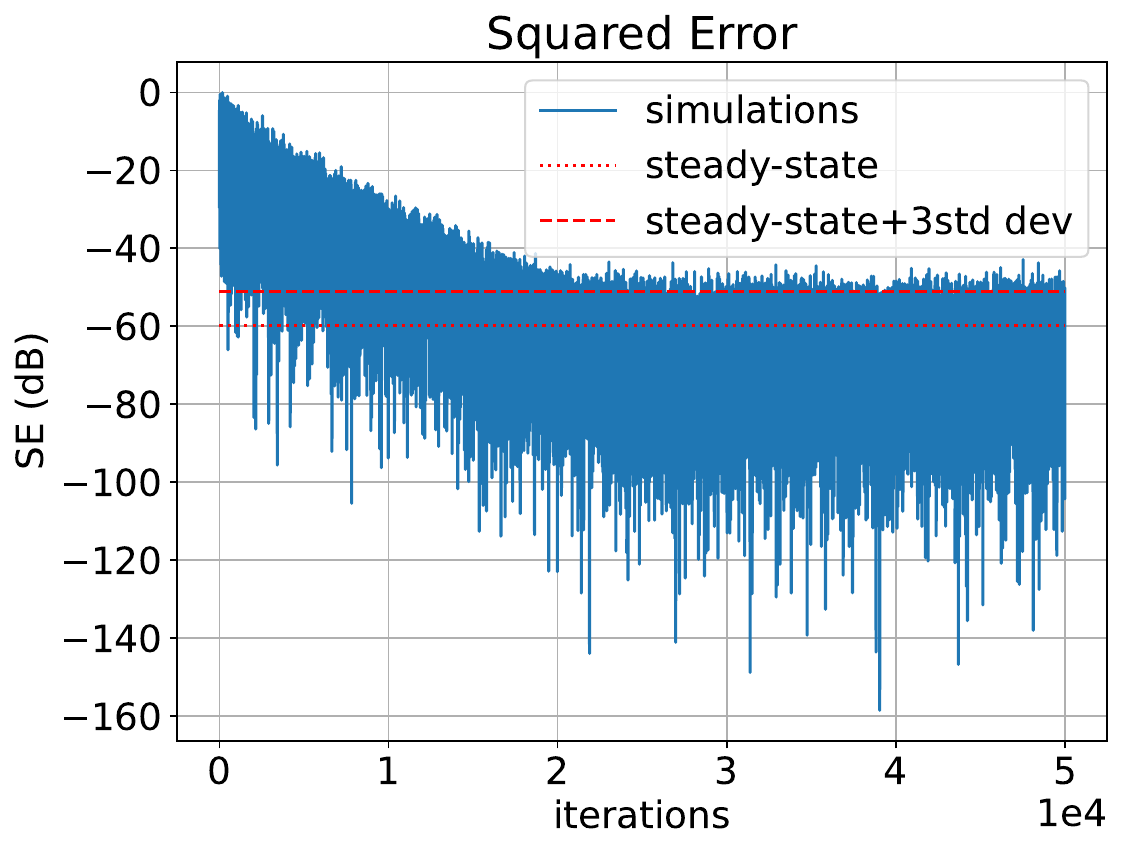}
        \caption{LMS (Laplacian noise)}
        \label{fig:LMSlaplacian}
     \end{subfigure}
     \begin{subfigure}[b]{.22\textwidth}
        \includegraphics[width=\textwidth]{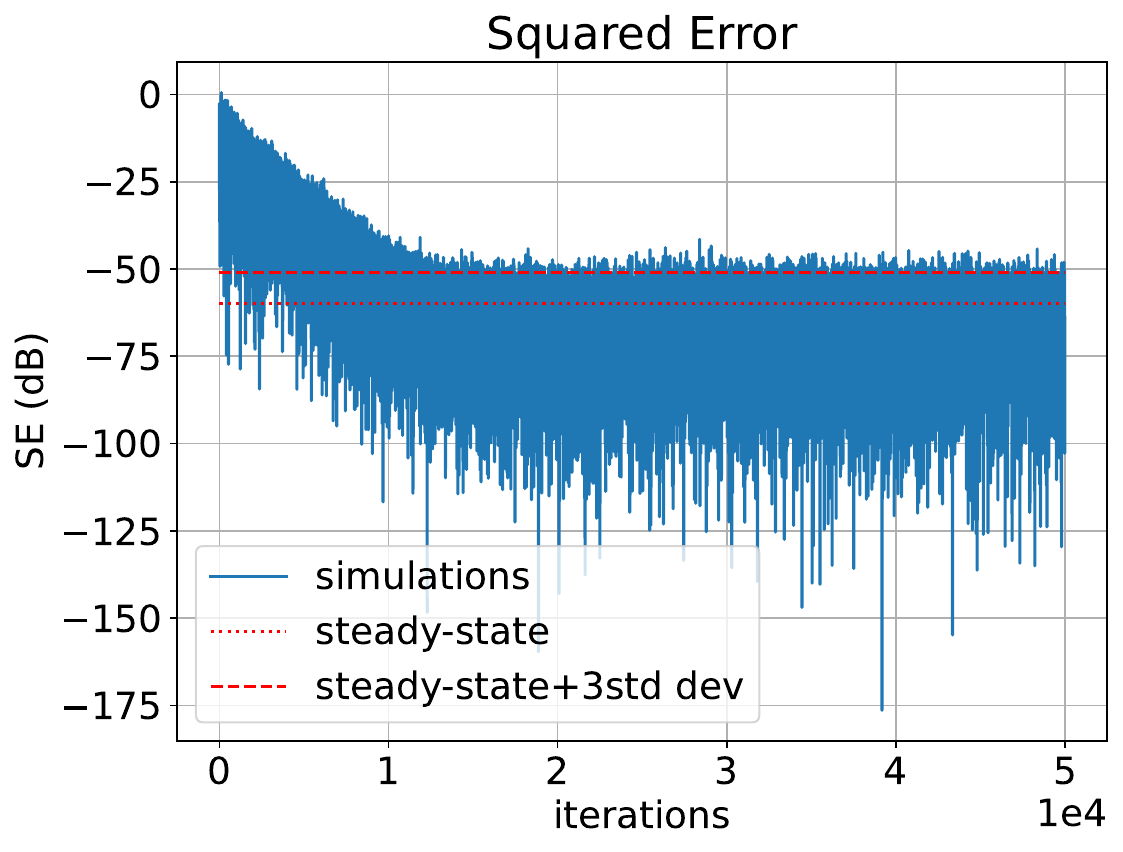}
        \caption{NLMS (Laplacian noise)}
        \label{fig:NLMSlaplacian}
     \end{subfigure}
     \hfill   
     \begin{subfigure}[b]{.22\textwidth}
        \includegraphics[width=\textwidth]{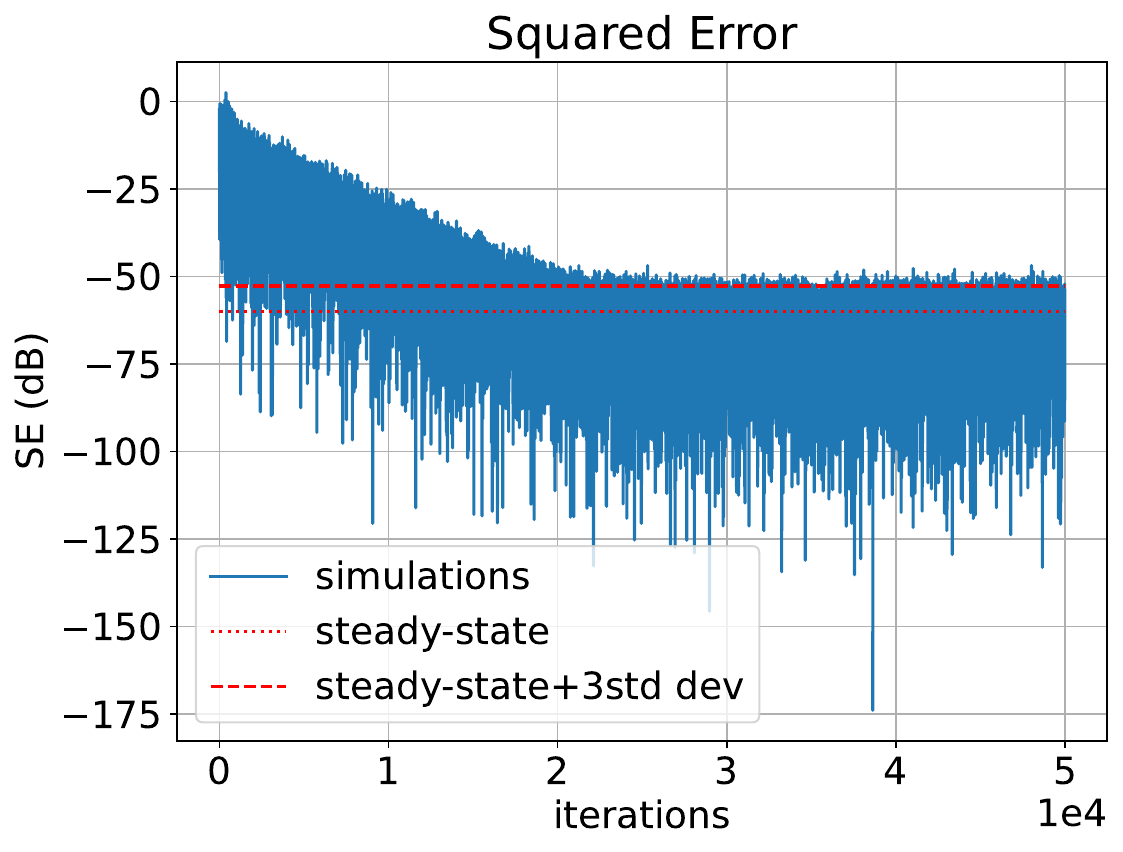}
        \caption{LMS (Gaussian power noise)}
        \label{fig:LMSewedian}
     \end{subfigure}
     \hfill   
     \begin{subfigure}[b]{.22\textwidth}
        \includegraphics[width=\textwidth]{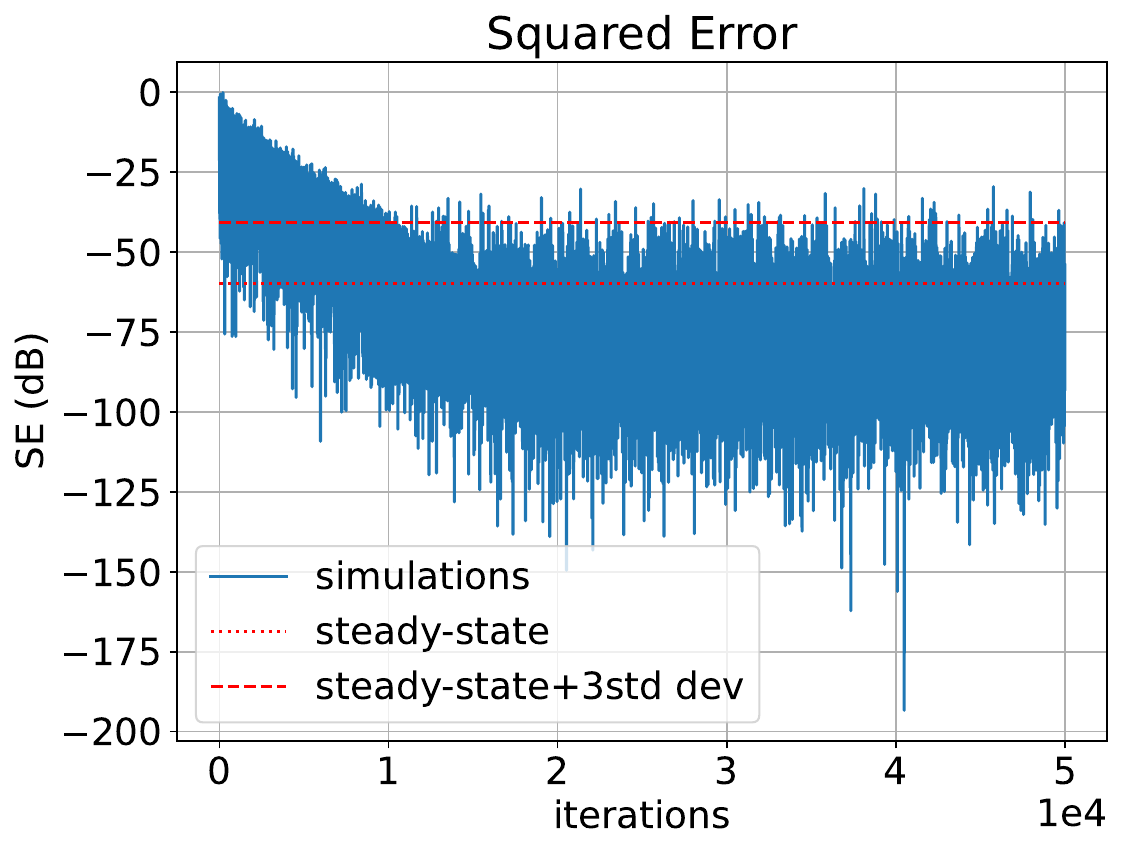}
        \caption{NLMS (Gaussian power noise)}
        \label{fig:NLMSewedian}
     \end{subfigure}
\caption{Sample curves for $\se^2[n]$ and predicted confidence intervals for non-Gaussian noises.}
\label{fig:nongaussian}
\end{figure}

Non-Gaussian zero-mean white noise with three different distributions with small, medium, and large kurtosis were considered. 1) uniform noise with $\kurtosisr=9/5$, 2) Laplacian noise with $\kurtosisr=6$, and 3) a Gaussian power noise $r[n] = u^5[n] \sqrt{\ssigmar^2/945}$ with $u[n] \sim {\cal N}(0,1)$, which has $\kurtosisr=733$~\cite{eweda_stochastic_2021}. The unknown system $\vh$ was the m5 response defined in~\cite{ITU-TG.168}. Figure~\ref{fig:nongaussian} shows the sample curves and confidence intervals empirically determined as $J[\infty]+3\sigma_{\se^2}$ using~\eqref{eqn:JinfLMS2} and~\eqref{eqn:maisfacildeinterpretar} (with $\mu =\beta/(\sN\sigma_\sx^2)$ for NLMS) for all noised distributions. These results show that the predictions using derived model are clearly useful for design purposes.

\section{Conclusion}

This letter has studied the behavior of a sample curve of the squared cancellation error for the LMS adaptive algorithm. Under reasonable simplifying assumptions, an analytical model has been derived for the  steady-state variance of the squared cancellation error as a function of the input and noise statistics.  Simulation results for Gaussian and non-Gaussian additive noises show that the predicted bounds for the squared error behavior can be very useful for design purposes. 

\bibliographystyle{IEEEtran}
\bibliography{references}

\end{document}